\documentclass[11pt,a4paper]{article}

\evensidemargin=\oddsidemargin \topmargin=0pt \advance\topmargin
by -\headheight \advance\topmargin by -\headsep

\begin{document}

\thispagestyle{empty} \quad

\vspace{2cm}
\begin{center}

{\Large \textbf{Heat kernel of non-minimal gauge field kinetic
operators on Moyal plane}}

\vspace{1.5cm}

{\large Alexei Strelchenko}
\\
{\large Dnepropetrovsk National University,\\
 49050
Dnepropetrovsk, Ukraine\\ E-mail: alexstrelch@yahoo.com }

\vspace{1.5cm}

{\large\textbf{Abstract}}

\end{center}

\begin{quote}
We generalize the Endo formula \cite{Endo:1984sz} originally
developed for the computation of the heat kernel asymptotic
expansion for non-minimal operators in commutative gauge theories
to the noncommutative case. In this way,  the first three non-zero
heat trace coefficients of the non-minimal  $U(N)$ gauge field
kinetic operator on the Moyal plane taken in an arbitrary
background are calculated. We show that the non-planar part of the
heat trace asymptotics is determined  by $U(1)$ sector of the
gauge model. The non-planar or mixed heat kernel coefficients are
shown to be gauge-fixing dependent in any dimension of space-time.
In the case of the degenerate deformation parameter the lowest
mixed coefficients in the heat expansion produce non-local
gauge-fixing dependent singularities of the one-loop effective
action that destroy the renormalizability of the $U(N)$ model at
one-loop level. Such phenomenon was observed at first in Ref.
\cite{Gayral:2004cu} for space-like noncommutative $\phi^{4}$
scalar and $U(1)$ gauge theories. The twisted-gauge transformation
approach is discussed.
\end{quote}
\vspace{1cm}

August 2006

\vspace{1cm}


\newpage

\section{Introduction}

The heat kernel of (pseudo)differential operators has become one
of the most powerful and actively developed tools
  in  quantum field theory and spectral geometry  (see
\cite{Gilkey:1994iy}, \cite{Berline:1992by},
\cite{Kirsten:2001ks}, \cite{Vassilevich:2003xt} where the
implementation of the heat kernel technique in  a variety of
physical and mathematical problems is discussed in details).
 Nowadays  this topic has acquired particular interest in the
context of noncommutative geometry and  quantum field theories on
noncommutative spaces \cite{Vassilevich:2003yz},
\cite{Gayral:2004ww}, \cite{Gayral:2004cs},
\cite{Vassilevich:2004ym}, \cite{Vassilevich:2005vk},
\cite{Gayral:2006vd}.  The main result here is that the
 heat trace
 for a differential operator on a (flat) noncommutative manifold, e.g. so-called generalized star-Laplacian arising, for
instance, in the noncommutative scalar $\lambda \varphi^{4}$
theory, can be expanded  in a power series in the "proper time"
parameter
   that resembles, in some respect,
  the heat trace expansion for its commutative
 counterpart. This observation is of fundamental importance
 since makes it possible to employ the heat kernel machinery in many applications to noncommutative models such as the investigation of one-loop
divergences or  quantum anomalies \cite{Vassilevich:2005vk}.

 Another interesting aspect of the heat kernel  on noncommutative spaces is closely
    related  to
  the UV/IR mixing phenomenon  \cite{Chepelev:1999ce}, \cite{Minwalla:1999we}, \cite{Arefeva:1999aa}.
  Namely, in the most general case when a star-differential operator involves both left and right Moyal multiplications
  (as it is for the generalized Laplacian mentioned above),
    its heat
 trace asymptotics   contains a contribution  produced by star-non-local terms that are singular when the deformation parameter
 vanishes. Clearly, it defines the non-planar part of the heat kernel expansion which is, in particular, responsible
 for the UV/IR mixing \cite{Gayral:2004cs},  \cite{Vassilevich:2005vk}. The situation gets even more
 intriguing in the case when the deformation parameter is
 degenerate (that corresponds to  space-like noncommutativity). In this case the non-planar contribution to the heat expansion becomes dangerous since
 it can
 affect the one-loop
 renormalization of a theory under consideration \cite{Gayral:2004cu}, \cite{Gayral:2006vd}.

In this paper we investigate the heat trace asymptotics for second
order star-differential operators\footnote{That is, we will
consider differential operators that contain star-products and
partial derivatives not higher then of second order. It should be
emphasized that such an operator
 is
 no longer a partial
differential operator  from the commutative point of view; even to
call it a pseudodifferential operator is not totally correct.
Indeed, the presence of the star-product,
  which is itself a differential
operator on the corresponding commutative manifold,  results in
the "incorrect" oscillatory behaviour of the symbol for the
  operator  and, hence, it cannot be
regarded as a pseudodifferential operator in the strict sense of
this term (see \cite{Vassilevich:2003yz} for the discussion of
this point).} on noncommutative non-compact flat manifolds without
boundary in the spirit of Ref. \cite{Vassilevich:2005vk}. We
restrict our consideration to the case of non-minimal operators
appearing in the non-commutative $U(N)$ gauge theory in the
background field formalism. To be precise, we are concerned
  with
   gauge field kinetic operator on the Moyal plane taken in the
covariant background gauge with an arbitrary gauge-fixing
parameter. In the commutative case
  non-minimal operators (in various physical systems)
 were investigated  by many
authors \cite{Endo:1984sz}, \cite{Barvinsky:1985bs},
\cite{Gusynin:1990ek}, \cite{Guendelman:1993ke},
\cite{Endo:1994yj}, \cite{Alexandrov:1996an},
\cite{Avramidi:2001tx}. In our study of the heat  asymptotics we
will follow the calculating method by Endo allowing to reduce the
whole task to the computation of the heat trace coefficients for
minimal operators by means of  some algebraic relations between
the heat kernel matrix elements \cite{Endo:1984sz},
\cite{Endo:1994yj}. Indeed, this method turns out to be especially
convenient  within the background field formalism; at the same
time, its purely algebraic nature allows one to generalize it
easily to the noncommutative case.

 The  paper
is organized  as follows.  The relevant basic formulae are briefly
reviewed in section 2. In section 3 we derive the non-commutative
version of the Endo formula   elaborated primarily for
computations in the commutative gauge theories.  In section 4 we
then calculate the heat trace coefficients of $U(1)$ gauge model;
the general case of $U(N)$ gauge symmetry is investigated in
details in Section 5. In section 6 we discuss the twisted gauge
transformation approach. Finally, we conclude with a summary
presented in section 7. To make the paper self-contained some
technical details on the evaluation of the heat kernel
coefficients are adduced in appendices.

 In the paper we adopt
the following conventions: small Greek letters from the beginning
of the alphabet, $~\alpha,~\beta, ~\gamma,~ \delta$, denote
indices of the $U(N)$
 inner group space; letters from the middle of the Greek alphabet, $~\lambda,~\mu,~\nu,
...$, refer to the indices of an $n$-dimensional Euclidean space.
Capital and small Latin letters are used to label generators of
the $U(N)$ and $SU(N)$ groups, respectively, i.e. $A,B,C=0,1,...,
N^{2}-1$ and $a,b,c=1,..., N^{2}-1$.

\section{Non-minimal operators in noncommutative\\ gauge theories}
Consider a self-adjoint   second order non-minimal
star-differential operator that corresponds to the  kinetic
operator of gauge particles propagating on Moyal plane in an
external background. It can be represented in the
form\footnote{Such an operator naturally appears in the NC $U(N)$
theory in the background field formalism; it defines, in
particular,   the quadratic in quantum gauge fields part of the
 total action written in a covariant background
gauge:
 $$S_{2}[Q]=-\frac{1}{2}\int_{R^n}d^{n}x
~\mathrm{tr}_{N}~ Q_{\mu}(x)D^{\xi}_{\mu\nu}Q_{\nu}(x),$$ where
$\mathrm{tr_{N}}$ means trace over internal indices (although we
do not write them explicitly) and  $Q_{\mu}$ describes quantum
fluctuations of the gauge fields. Functional integration  of the
expression $\exp{S_{2}[Q]}$, as known, gives the one-loop
effective action, $\Gamma_{gauge}[B]= \frac{1}{2}\ln
\det(D^{\xi})$, that is invariant under the background field
(star)gauge transformations of the form $\delta
B_{\mu}(x)=\nabla_{\mu}\lambda(x)$.
 Some aspects of the background
field formalism in NC field theories can be found, for instance,
in Ref. \cite{Das}.}
\begin{equation}\label{one}
D^{\xi}_{\mu\nu}=-\Bigl[\delta_{\mu\nu}\nabla^{2}+(\frac{1}{\xi}-1)\nabla_{\mu}
 \nabla_{\nu} +2( L(F_{\mu\nu})-R(F_{\mu\nu}))\Bigr],
\end{equation}
where  $$\nabla_{\mu}=\partial_{\mu}+ L( B_{\mu})- R(B_{\mu})$$ is
an anti-Hermitian covariant-derivative operator in the background
field
 $B_{\mu}$,
 $\xi$ is a numerical
gauge-fixing parameter and $F_{\mu\nu}$ is the curvature tensor of
the gauge connection $B_{\mu}$. Here operator $L$ (accordingly,
$R$) involves both left (right) Moyal  and left (right) matrix
multiplications, i. e.
 $$L(l)f=l \star f, ~~~ R(r)f=f\star
r,$$ with $f,~l$ and $r$ being matrix valued functions; the Moyal
star-product on $R^{n}$ can be defined as  $$l(x) \star
f(x)=l(x)\exp{\left(\frac{\imath}{2}\theta^{\mu\nu}\overleftarrow{\partial}_{\mu}\overrightarrow{\partial}_{\nu}\right)}f(x),$$
where $\theta$ is a constant antisymmetric matrix (in practice it
is sometimes convenient to use the Reiffel representation of the
star-product, see Appendix B).
 Commutator of the
covariant derivatives gives
\begin{eqnarray*}\label{two}
 [\nabla_{\mu},\nabla_{\nu}]=  L(F_{\mu\nu})- R(F_{\mu\nu}),~~~
F_{\mu\nu}=\partial_{\mu}B_{\nu}-\partial_{\nu}B_{\mu}+[B_{\mu},B_{\nu}]_{\star}.
\end{eqnarray*}
  We denote the operator
$-\nabla^{2}$
 as $  D_{0}$ which is a self-adjoint non-negative operator
 corresponding to the inverse propagator of ghost particles.
In the following we assume that the operators $D_0$ and
$D^{\xi}_{\mu\nu}$ have no zero-modes.

To simplify  our
 analysis let us consider the  case of $U(1)$ gauge symmetry
(generalization to the case of $U(N)$ symmetry will be discussed
in Section 5). The heat trace for the kinetic operator (\ref{one})
is defined as
\begin{equation}\label{three}
  K^{\xi}(t)=Tr_{L^{2}}\exp(-tD^{\xi}),
\end{equation}
 where $t$
is a (positive) spectral parameter and the trace is taken on the
space of square integrable functions \cite{Kirsten:2001ks},
\cite{Vassilevich:2003xt}. Usually this expression is regularized
by subtracting the heat trace of the Laplacian
$\triangle=-\partial_{\mu}\partial^{\mu}$ since the small $t$
asymptotic expansion of the quantity  $Tr_{L^{2}}\exp(-tD^{\xi})$
contains a volume term that is divergent on a non-compact
manifold.

 We wish to compute the heat trace in the limit of small
spectral parameter $t \rightarrow 0$ by means of the
 Fock-Schwinger-DeWitt proper-time
method.  To this aim we introduce two abstract Hilbert spaces
spanned by basis vectors $|x \rangle$ and $|\mu,x \rangle$,
respectively, and define "Hamiltonian" operators $\widehat{D}_{0}$
and $\widehat{D}^{\xi}$ associated with  $D_{0}$ and
$D^{\xi}_{\mu\nu}$ by\footnote{In this paper we are dealing with
flat Euclidean space and therefore the distinction between upper
and lower indices is irrelevant.}
\begin{eqnarray}\label{four}
\langle x|\widehat{D}_{0}|x' \rangle=D_{0}\langle x|x'
\rangle,~~~~~~~~~~~~~ \nonumber\\
 \langle x, \mu|\widehat{D}^{\xi}|\nu,x' \rangle=D^{\xi}_{\mu\lambda}\langle x,
 \lambda|\nu,x'
 \rangle.
\end{eqnarray}
Operators on the right hand sides of these expressions are viewed
as differential operators with respect to the variable $x$. The
basis vectors satisfy  the orthonormality conditions
 $$\langle x|x' \rangle =\delta(x,x'),~~~~~~~~~$$
 $$ \langle x, \mu|\nu,x' \rangle =\delta_{\mu\nu}\delta(x,x').$$
Note that, in the case of an arbitrary manifold, index of $|\mu,x
\rangle$ (as well as that of the conjugate $\langle x, \mu|$) is
regarded as that of a covariant vector density of weight $1/2$
\cite{DeWitt:1965dt}.

Next, the proper-time transformation functions, or heat kernels,
 for the operators $D_{0}$ and
$D^{\xi}_{\mu\nu}$ are introduced  by
\begin{eqnarray}\label{five}
K_{0}(x, x'; t)= \langle x|\exp[-t \widehat{D}_{0}]|x'
\rangle,~~~~~~~ \nonumber
\\ K^{\xi}_{\mu\nu}(x, x'; t)= \langle x, \mu|\exp[-t
\widehat{D}^{\xi}]|\nu,x' \rangle,
\end{eqnarray}
where $t$ is interpreted as the proper-time
parameter\footnote{That is, the exponential operators on the right
hand sides of (\ref{five}) can be regarded as evolution operators
of a "particle" in the proper time $t$ \cite{Schwinger:1951sr}.}.
By making use of (\ref{four}) it can be straightforwardly checked
that the kernels $K_{0}(x, x'; t)$ and $K^{\xi}_{\mu\nu}(x, x';
t)$ satisfy the heat equations:
\begin{eqnarray}\label{six}
\left( \frac{\partial}{\partial t}+D_{0}\right)K_{0}(x, x';
  t)=0,~~~~~~~ \nonumber \\
 \left( \delta_{\mu \lambda}\frac{\partial}{\partial t}+D^{\xi}_{\mu \lambda}\right)K^{\xi}_{\lambda\nu}(x, x';
  t)=0,
\end{eqnarray}
with the boundary conditions
\begin{eqnarray}\label{seven}
 \lim_{t \rightarrow 0}K_{0}(x, x'; t)=\delta(x,x'), ~~~~
  \lim_{t \rightarrow 0}K^{\xi}_{\mu\nu}(x, x';
  t)=\delta_{\mu\nu}\delta(x,x').
\end{eqnarray}

 From the kernels (\ref{five}) one can obtain the one-loop  effective action of pure NC Yang-Mills theory using the
standard formal expressions:
\begin{eqnarray}\label{nine}
  \Gamma^{(1)}[B]=\Gamma_{gauge}[B]+\Gamma_{ghost}[B], ~~~~~~~~~~~~~~~~~~~~\nonumber\\
 \Gamma_{gauge}[B]= \frac{1}{2}\ln \det(D^{\xi})=-\frac{1}{2}\int_{R^{n}}dx\int_{0}^{\infty}\frac{dt}{t}~tr_{V} K^{\xi}_{\mu\nu}(x, x;
  t),\\
 \Gamma_{ghost}[B]=-
\ln\det(D)=\int_{R^{n}}dx\int_{0}^{\infty}\frac{dt}{t}~tr_{V}
K_{0}(x, x;
  t),~~~~\nonumber
\end{eqnarray}
where the first term, $\Gamma_{gauge}[B]$, describes a
contribution to the effective action  coming from the gauge sector
of the model while the second term, $\Gamma_{ghost}[B]$, stands
for the ghost contribution; $tr_{V}$ means trace over Euclidean
vector and, in general, internal  indices.
 As  is well-known, the expression
for  $\Gamma^{(1)}[B]$ is divergent and must be regularized. This
can be done, for instance, by  replacing $1/t$ in the integrands
of (\ref{nine}) with $\mu^{2\epsilon}/t^{1-\epsilon}$, where
$\epsilon$ is a complex parameter and $\mu$ is a dimensional
quantity introduced to keep the total mass dimension of the
expression unchanged.  Now all information on the one-loop
effective action contains in the heat traces which at $t
\rightarrow 0^{+}$ can be expanded in series over the spectral
(proper time) parameter :
\begin{equation}\label{ninea}
Tr K(D;
  t)\simeq \sum_{k=0}^{\infty}t^{(k-n)/2}a_{k}(D).
\end{equation}
The coefficients $a_{k}(D)$ here define the asymptotics of the
heat trace as $t \rightarrow 0$. On the manifold without boundary
odd-numbered coefficients are equal to zero. From the expressions
(\ref{nine}) and (\ref{ninea}) one sees that terms with $k \leq n$
in the heat kernel expansion can potentially  give rise to
divergences in the effective action (see also discussion in the
end of section 5).

In the commutative case the heat kernel coefficients $a_{k}$,
known also as diagonal Seeley-Gilkey-DeWitt coefficients
 \cite{DeWitt:1965dt},
\cite{Seeley:1967sl}, \cite{Gilkey:1975ge}, are expressed only in
terms of local gauge covariant quantities, such as matter fields,
gauge field strength tensor and their covariant
derivatives\footnote{At finite temperature there is a further
gauge covariant quantity, that is the (untraced)  Polyakov loop
(see Ref. \cite{Megias:2003ui} and references therein).}, and,
hence, are manifestly gauge invariant objects
 (see review
article \cite{Vassilevich:2003xt}). However, on $\theta$-deformed
manifolds, there appears another type of coefficients in the heat
kernel expansion (\ref{ninea}), so-called mixed coefficients, that
reflect the non-local nature of  NC field theories
\cite{Gayral:2004cs}, \cite{Vassilevich:2005vk}. As we have
mentioned earlier, the contribution of these mixed terms is
equivalent to the contribution of non-planar diagrams to the
effective action. In particular, it can develop non-local
singularities as $\epsilon \rightarrow 0$ if the deformation
parameter is degenerate \cite{Gayral:2004cu}. We will comment this
point later on.

\section{Noncommutative Endo formula}
Consider the heat trace for the operator (\ref{one}), $Tr
K^{\xi}_{\mu\nu}(x, x;
  t)$, and compute the first three non-zero heat kernel coefficients in the small $t$ asymptotic  expansion for this quantity.
   In the Feynman gauge ($\xi=1$) it can be done by means of the
calculating procedure described in Refs.
\cite{Vassilevich:2003yz}, \cite{Gayral:2004ww},
\cite{Vassilevich:2005vk}. To apply it in the more general case of
an arbitrary value of the gauge-fixing parameter we will reproduce
in what follows the non-commutative version of the Endo formula
\cite{Endo:1984sz}, \cite{Endo:1994yj}. To simplify our
computations we suppose that the background field satisfies the
equation of motion:
\begin{equation}\label{nineb}
\nabla_{\mu}F_{\mu\nu}=\partial_{\mu}F_{\mu\nu}+[B_{\mu},
F_{\mu\nu}]_{\star}=0.
\end{equation}
 First, we note the obvious relation
\begin{equation}\label{ten}
[\nabla_{\mu}, \nabla^{2}]=2  \{L(F_{\mu\nu})-
R(F_{\mu\nu})\}\nabla_{\nu}.
\end{equation}
From (\ref{one}), (\ref{nineb}) and (\ref{ten}) it is easily seen
that
\begin{equation}\label{eleven}
\nabla_{\mu}\nabla_{\lambda}D^{\xi}_{\lambda\nu}=D^{\xi}_{\mu\lambda}\nabla_{\lambda}\nabla_{\nu}=-\frac{1}{\xi}\nabla_{\mu}\nabla^{2}\nabla_{\nu}
\end{equation}
and
\begin{equation}\label{twelv}
\nabla_{\mu}\nabla_{\lambda}D^{\xi}_{\lambda\nu}=\frac{1}{\xi}\nabla_{\mu}\nabla_{\lambda}D^{(\xi=1)}_{\lambda\nu}.
\end{equation}
The first equality in (\ref{eleven}) means that the operators
 $\nabla_{\mu}\nabla_{\lambda}$ and $D^{\xi}_{\lambda\nu}$
commute each other, as they do in the commutative case. With the
help of (\ref{eleven}) and (\ref{twelv}) one arrives at
\begin{equation}\label{thirteen}
\nabla_{\mu}\nabla_{\lambda}\Bigl[D^{\xi~n}\Bigr]_{\lambda\nu}=\frac{1}{\xi^{n}}\nabla_{\mu}\nabla_{\lambda}\Bigl[\overline{D}^{~n}\Bigr]_{\lambda\nu},
\end{equation}
where we denote $\overline{D}_{\mu\nu}=D^{(\xi=1)}_{\mu\nu}$, and
the expression $\Bigl[D^{~n}\Bigr]_{\mu\nu}$ stands for the
$n^{th}$ power of the operator $D_{\mu\nu}$ in the usual
sense\footnote{Remind that the operators $D^{(\xi)}$ and
$\overline{D}$ contain  both left and right star-multiplications.
}, i.e.
$$\Bigl[D^{\xi~n}\Bigr]_{\mu\nu}=D^{\xi}_{\mu\lambda_1}D^{\xi}_{\lambda_1\lambda_2}\ldots
D^{\xi}_{\lambda_{n-1}\nu}.$$ In particular, with the help of
(\ref{twelv}), one gets the following useful relation
\begin{equation}\label{fourteen}
\nabla_{\mu}\nabla_{\lambda}\exp(-t
D^{\xi})_{\lambda\nu}=\nabla_{\mu}\nabla_{\lambda}\exp\left(-\frac{t}{\xi}~
\overline{D}\right)_{\lambda\nu}.
\end{equation}
With these expressions we are ready to derive the non-commutative
Endo formula. For this purpose we differentiate both sides of
(\ref{five}) with respect to $1/\xi$ to get:
\begin{equation}\label{fifteen}
 \frac{\partial}{\partial\xi^{-1}}K^{\xi}_{\mu\nu}(x, x';
  t)=\nabla_{\mu}\nabla_{\lambda}K^{\xi}_{\lambda\nu}(x, x';
  t)=t~\nabla_{\mu}\nabla_{\lambda}\overline{K}_{\lambda\nu}(x, x';
  \frac{t}{\xi}),
\end{equation}
and then integrate the obtained relation over $\xi^{-1}$. Here
$\overline{K}_{\mu\nu}(t)=K^{\xi=1}_{\mu\nu}(t)$.  After
redefinition of the integration parameter one arrives  at (cf.
expr. (2.23) in Ref. \cite{Endo:1984sz})
\begin{equation}\label{sixteen}
K^{\xi}_{\mu\nu}(x, x';
  t)=\overline{K}_{\mu\nu}(x, x';
  t)+\int_{t}^{\frac{t}{\xi}}d\tau~\nabla_{\mu}\nabla_{\lambda}\overline{K}_{\lambda\nu}(x, x';
  \tau).
\end{equation}
It is easy to show that this expression satisfies the heat kernel
equation (\ref{six}) as it should. The main advantage of the
formula (\ref{sixteen}) is that one deals now with the kernel of
the minimal operator $D^{\xi=1}_{\mu\nu}$, i.e.
$\overline{K}_{\mu\nu}(x, x';t)$, which is much more convenient in
practical computations.

Equation (\ref{sixteen}) can be simplified further. To do this we
need the Ward identity for the heat kernels which is expressed by
the relation\footnote{The commutative analogue of this expression
is derived in Refs. \cite{Endo:1984sz}, \cite{Barvinsky:1985bs},
\cite{Nielsen:1988nl}.}:
\begin{equation}\label{eighteen}
\nabla_{\lambda} K^{\xi}_{\lambda\nu}(x, x';
  \tau)=-\nabla'_{\nu}K_{0}\left(x, x';
  \frac{\tau}{\xi}\right).
\end{equation}
Here and further on prime over nabla on RHS of (\ref{eighteen})
indicates that covariant derivative acts on $x'$ variable.

To prove (\ref{eighteen})  we note  that for the operators of the
left and right Moyal multiplications there are the following
rules: $$\langle x, \mu|\widehat{L}(B)|\nu,x'
\rangle:=L(B(x))\langle x, \mu|\nu,x' \rangle=R(B(x'))\langle x,
\mu|\nu,x' \rangle$$ and $$\langle x, \mu|\widehat{R}(B)|\nu,x'
\rangle:=R(B(x))\langle x, \mu|\nu,x' \rangle=L(B(x'))\langle x,
\mu|\nu,x' \rangle.$$ As a consequence, one has
$$\nabla_{\lambda}\langle x, \mu|\nu,x'
\rangle=-\nabla'_{\lambda}\langle x, \mu|\nu,x' \rangle,$$ that
is, at $\tau=0$ left- and right-hand sides of the Ward identity
(\ref{eighteen}) agree by the heat kernel boundary conditions
(\ref{seven}).

Next, following Endo, we define an operator $\widehat{\delta}$
that connects both Hilbert spaces through the relation: $$\langle
x|\widehat{\delta}|\nu,x' \rangle=-\nabla_{\mu}\langle x,
\mu|\nu,x' \rangle=\nabla'_{\nu}\langle x~|x'~ \rangle.$$ From
(\ref{eleven}) one can see that
$$\nabla_{\lambda}D^{\xi}_{\lambda\nu}=\frac{1}{\xi}D_{0}\nabla_{\nu},$$
 which is written in the operator form  as
$$\widehat{\delta}~\widehat{D}^{\xi}=\frac{1}{\xi}~\widehat{D}_{0}~\widehat{\delta}.$$
By making use of the definition for the operator
$\widehat{\delta}$,
 one obtains the needed identity:
\begin{eqnarray}\label{seventeen}
\nabla_{\lambda}K^{\xi}_{\lambda\nu}(x, x';
  \tau)=\nabla_{\lambda}\langle x,
  \lambda|\exp(-\tau~\widehat{D}^{\xi})|\nu,x'
  \rangle =-\langle
  x|\widehat{\delta}\exp(-\tau~\widehat{D}^{\xi})|\nu,x'
  \rangle ~~~~~\nonumber\\
 =-\langle x|\exp\left(-\frac{\tau}{\xi}~\widehat{D}_{0}\right)\widehat{\delta}|\nu,x'
  \rangle=-\nabla'_{\nu}K_{0}\left(x, x';
  \frac{\tau}{\xi}\right).~~~~
\end{eqnarray}
In particular, for $\xi=1$ we have:
\begin{equation}\label{eighteenprime}
\nabla_{\lambda}\overline{K}_{\lambda\nu}(x, x';
  \tau)=-\nabla'_{\nu}K_{0}(x, x';
  \tau),
\end{equation}
and the expression (\ref{sixteen}) can be represented now in the
form
\begin{equation}\label{nineteen}
K^{\xi}_{\mu\nu}(x, x';
  t)=\overline{K}_{\mu\nu}(x, x';
  t)-\int_{t}^{\frac{t}{\xi}}d\tau~\nabla_{\mu}\nabla'_{\nu}K_{0}(x, x';
  \tau).
\end{equation}
Formula (\ref{nineteen}) is the starting point of our
computations. More precisely, we are going to investigate the heat
 asymptotics for the trace of the kernel (\ref{nineteen}). In this
 connection it is necessary to note that the operators $\exp{(-t
 D^{\xi=1})}$ and $\exp{(-t
 D_{0})}$ are  trace-class for positive  values of the spectral parameter $t$ and,
 hence, the asymptotic expansions  for the kernels in RHS of
 (\ref{nineteen}) are well-defined \cite{Vassilevich:2003yz},
\cite{Gayral:2004ww}, \cite{Vassilevich:2005vk},
\cite{Gayral:2006vd}.

\section{Evaluation of the heat kernel coefficients}
 We wish to calculate the  heat kernel coefficients in the asymptotic expansion for the quantity $Tr
K^{\xi}_{\mu\nu}(t)$. In what follows we assume that the
$\theta$-parameter
 is non-degenerate\footnote{Hence we are working in
even-dimensional manifolds. The case of the degenerate  parameter
$\theta$ will be discussed in the end of the next section. }.

 The asymptotic expansion for the
first term on RHS of (\ref{nineteen})  is, in fact, investigated
in
 Ref. \cite{Vassilevich:2005vk} where the heat kernel coefficients for generalized Laplacians on the Moyal plane
containing both left and right multiplications were calculated.
The result is presented in (\ref{thirty}) below. Consider the
second term. After performing an integration by parts and taking
the cyclic property of the Moyal product into account one gets
\begin{eqnarray}\label{twentyone}
  Tr  \nabla_{\mu}\nabla'_{\nu}K_{0}(\tau)=\int_{R^{n}}dx ~~ \Bigl[\nabla_{\mu}e^{-t D_{0}} \langle x~|~ x'\rangle \overleftarrow{\nabla}'_{\mu}  \Bigr]_{x=x'}~~~~~~~~~~~~~~~~~~\nonumber \\
  =-\int_{R^{n}}dx ~~   \Bigl[\nabla_{\mu} \nabla_{\mu}e^{-t D_{0}}\langle x~|~ x'\rangle \Bigr]_{x=x'}.~~~~~~~~~~~~~~~
\end{eqnarray}
As it was mentioned in the end of the previous section,  for this
expression there is an  asymptotic expansion
\begin{equation}\label{twentytwo}
 Tr  \nabla_{\mu}\nabla'_{\nu}K_{0}(\tau)\simeq
\sum_{k=-2}^{\infty}t^{(k-n)/2}\widetilde{a}_k(\nabla^{2}, D_0),
\end{equation}
 where the coefficients\footnote{We use
tilde to indicate that these coefficients correspond to the
quantity $Tr  \nabla_{\mu}\nabla'_{\nu}K_{0}(\tau)$. In the end we
have to carry out integration over the spectral parameter in the
expression (\ref{twentytwo}), in accordance with the formula
(\ref{twentyeight}) below.} $\widetilde{a}_k(\nabla^{2}, D_0)$ can
be decomposed
 as ( cf. \cite{Vassilevich:2005vk})
\begin{equation}\label{twentythree}
  \widetilde{a}_k(\nabla^{2}, D_0)=\widetilde{a}^{planar}_k(\nabla^{2}, D_0)+\widetilde{a}^{mixed}_k(\nabla^{2},
  D_0).
\end{equation}
 Here the coefficients $\widetilde{a}^{planar}_k(\nabla^{2}, D_0)$
are expressed as integrals of gauge invariant star polynomials of
the fields (i.e. with the deformation parameter being hidden in
the Moyal products). These coefficients contribute to the planar
part of the heat expansion. The other type of the heat kernel
coefficients, $\widetilde{a}^{mixed}_k(\nabla^{2}, D_0)$,
corresponds to the contributions from non-planar diagrams in the
diagrammatic language; these are so-called mixed heat kernel
coefficients.

There are several ways to compute the planar heat kernel
coefficients. One of them is based on the functorial properties of
the heat kernel described in \cite{Gilkey:1975ge},
\cite{Branson:1997Bn}, \cite{Vassilevich:2003xt}.  To this aim one
has to re-express (\ref{twentyone}) in the form\footnote{One has
to expand formally the operator exponent in the integrand of
(\ref{twentyone}) into power series over the spectral parameter
and single out terms which contain only left and right Moyal
multiplications, respectively.}
\begin{eqnarray}\label{twentyfour}
 \int_{R^{n}}dx ~~   \Bigl[\nabla_{\mu} \nabla_{\mu}e^{-t D_{0}}\langle x~|~ x'\rangle \Bigr]_{x=x'}
 =~~~~~~~~~~~~~~~~~~~~~~~~\nonumber\\
= \int_{R^{n}}dx ~   \Bigl[( \nabla^{L}_{\mu}
\nabla^{L}_{\mu}e^{-t D^{L}_{0}}+\nabla^{R}_{\mu}
\nabla^{R}_{\mu}e^{-t D^{R}_{0}}~~~~~\nonumber\\
~~~~~~~~~~~~~~~~~-\partial^{2}e^{t\partial^{2}}+mixed~
terms)\langle x~|~ x'\rangle \Bigr]_{x=x'},
\end{eqnarray}
where $$\nabla^{L}_{\mu}=\partial_{\mu}+L(B_{\mu}),~~~
D^{L}_{0}:=-\nabla^{L}_{\mu}\nabla^{L}_{\mu},$$ and
$$\nabla^{R}_{\mu}=\partial_{\mu}-R(B_{\mu}),~~~
D^{R}_{0}:=-\nabla^{R}_{\mu}\nabla^{R}_{\mu}.$$
 The first and the
second terms of the integrand contain only left and right Moyal
multiplications in all operators, respectively, and produce planar
coefficients in the expansion (\ref{twentytwo}). These terms can
be considered as a non-commutative analogue of vacuum expectation
values of the second order differential operators with scalar
leading symbol \cite{Branson:1997Bn}:
 $$\langle
D^{L,R}_{0}\rangle:=Tr_{L^{2}}D^{L,R}_{0} \exp{(-t
D^{L,R}_{0})}.$$ For such terms one has to apply  result of the
Theorem 3.3 of Ref. \cite{Branson:1997Bn} generalized to the
non-commutative case (remind that here we are concerned with
trivial metric of the flat space). For the sake of completeness we
present some of the necessary details in Appendix A.
  The third term in RHS of (\ref{twentyfour}) is needed to kill an extra volume term while the last one
 stands for the contribution of the mixed terms and
must be studied separately.

More straightforward method to obtain the coefficients
(\ref{twentythree}) consists in making use of a particular basis
in the Hilbert space of square integrable functions on
$R^{n}_{\theta}$. The advantage of this method is that it can be
employed  for evaluation of the mixed heat kernel coefficients as
well. As basis vectors $|x~\rangle$ we take plane waves. After
some simple manipulations one obtains
\begin{eqnarray}\label{twentyfive}
  Tr  \nabla_{\mu}\nabla'_{\nu}K_{0}(\tau)=\int_{R^{n}}dx \int \frac{d^{n}k}{(2\pi)^{n}}\nabla_{\mu}e^{-t D_{0}} e^{ik
  x}\star \nabla_{\mu} e^{-ik
  x}~~~~~~~~~~~~~~~~~~~~~\nonumber \\
  =-\int_{R^{n}}dx \int \frac{d^{n}k}{(2\pi)^{n}}  e^{-ik
  x} \star \nabla_{\mu} \nabla_{\mu}e^{-t D_{0}} e^{ik
  x} =-\int_{R^{n}}dx \int \frac{d^{n}k}{(2\pi)^{n}}e^{-t k^{2}}\nonumber\\
 \times e^{-\imath kx}\star \nabla_{\mu}\nabla_{\mu}
  \exp{[t \{(\nabla-\imath k)^{2}+2 \imath k_{\mu}(\nabla_{\mu}-\imath k_{\mu})\}]}e^{\imath k
  x}.
\end{eqnarray}
  Next, by expanding the exponential $\exp{[t (\nabla-\imath k)^{2}+2 \imath k_{\mu}(\nabla_{\mu}- \imath k_{\mu})]}$ in a power
series in $(\nabla_{\mu}- \imath k_{\mu})$
 and making use of the calculating procedure presented in Refs. \cite{Vassilevich:2003yz}, \cite{Vassilevich:2005vk}
 (see also original paper \cite{Nepomechie:1985nm}) one gets for the first two  planar heat kernel coefficients:
\begin{eqnarray}\label{twentysix}
  \widetilde{a}^{planar}_{2}(\nabla^{2}, D_0)&=&\frac{1}{(4 \pi)^{\frac{n}{2}}}\int_{R^{n}}dx
  \frac{4-n}{24}F_{\mu\nu}\star
F_{\mu\nu},\nonumber\\
 \widetilde{a}^{planar}_{4}(\nabla^{2}, D_0)&=&\frac{1}{(4
 \pi)^{\frac{n}{2}}}\int_{R^{n}}dx\frac{n-6}{360}(6 F_{\mu\nu}\star
F_{\nu\lambda}\star F_{\lambda\mu}\\
&~&~~~~~~~~~~+2\nabla_{\mu}F_{\nu\lambda}\star\nabla_{\mu}F_{\nu\lambda}-\nabla_{\mu}F_{\mu\lambda}\star\nabla_{\nu}F_{\nu\lambda})\nonumber.
\end{eqnarray}
Notice that the first coefficient of the expansion
(\ref{twentytwo}), $ \widetilde{a}^{planar}_{-2}$, represents a
field independent  volume divergence and is not indicated here. It
can be absorbed, for instance, by subtracting the quantity
$Tr_{L^{2}}\triangle e^{-t\triangle}$, where
$\triangle=-\partial_{\mu}\partial^{\mu}$ (see the  remark to Eq.
(\ref{three})). In the following we will always omit such volume
terms. It should be emphasized that the background field must
satisfy certain fall-off condition to secure the convergence of
the integrals in (\ref{twentysix}).
 We remark also
that $\widetilde{a}^{planar}_{0}(\nabla^{2}, D_0)=0$ as it can be
expected since there is no gauge-invariant object corresponding to
gauge-field mass term in the effective action. The first non-zero
mixed coefficient reads \cite{Vassilevich:2005vk} (see Appendix B
for details):
\begin{eqnarray}\label{twentyseven}
  \widetilde{a}^{mixed}_{n}(\nabla^{2}, D_0)=-2 (det~\theta)^{-1}(2\pi)^{-n}\int_{R^{n}}dx
\int_{R^{n}}dy B_{\mu}(x) B_{\mu}(y).
\end{eqnarray}
 Finally, to obtain the heat trace coefficients for the operator (\ref{one}) one has to substitute these expressions to the traced formula
(\ref{nineteen}),
\begin{equation}\label{twentyeight}
Tr K^{\xi}_{\mu\nu}(t)=Tr \overline{K}_{\mu\nu}(t)-Tr
\int_{t}^{\frac{t}{\xi}}d\tau~\nabla_{\mu}\nabla'_{\nu}K_{0}(\tau),
\end{equation}
 and carry out integration over parameter $\tau$.
The heat kernel coefficients for the minimal operator
$D^{\xi=1}_{\mu\nu}$ read \cite{Gayral:2004ww},
\cite{Vassilevich:2005vk}
\begin{eqnarray}\label{twentynine}
  a^{planar}_{4}&=&\frac{1}{(4 \pi)^{\frac{n}{2}}}\int_{R^{n}}dx
  \left(\frac{n}{6}-1\right)F_{\mu\nu}\star
F_{\mu\nu},\\
 a^{planar}_{6}&=&\frac{1}{(4
\pi)^{\frac{n}{2}}}\frac{1}{360}\int_{R^{n}}dx [120
   F_{\mu\nu}\star
F_{\nu\lambda}\star F_{\lambda\mu} -60F_{\mu\nu}\star\nabla^{2}
F_{\mu\nu}\nonumber\\ &-& 2n(6 F_{\mu\nu}\star F_{\nu\lambda}\star
F_{\lambda\mu}+2\nabla_{\mu}F_{\nu\lambda}\star\nabla_{\mu}F_{\nu\lambda}-\nabla_{\mu}F_{\mu\lambda}\star\nabla_{\nu}F_{\nu\lambda})\nonumber
\end{eqnarray}
for the first two non-trivial planar coefficients (remind that we
neglect a trivial volume term) and
\begin{eqnarray}\label{thirty}
a^{mixed}_{n+2}&=&(det~\theta)^{-1}\frac{2n}{(2\pi)^{n}}\int_{R^{n}}dx
\int_{R^{n}}dy B_{\mu}(x) B_{\mu}(y)
\end{eqnarray}
for the first mixed one. The coefficients (\ref{twentysix}),
(\ref{twentyseven}), (\ref{twentynine}), (\ref{thirty}) are gauge
invariant as it should be. In the next section we will investigate
the general case of $U(N)$ gauge symmetry. In particular, we will
find that the corresponding mixed coefficients are determined only
by $U(1)$ sector of the gauge model.

\section{U(N) gauge symmetry}

  Let $T^{A}$, $A=0,1,...,N^2-1$, be the
generators of the $U(N)$ group in the fundamental representation.
The background potential is represented as
$B_{\mu}=B^{A}_{\mu}T^{A}$ that is a $N \times N$ matrix in the
group space.
 We normalize the  $U(1)$ generator as follows
$T^{0}=\frac{1}{\sqrt{2N}}$, so that $$\mathrm{tr_{N}}~
T^{A}T^{B}=\frac{1}{2}\delta^{AB}.$$  The generators of the
$SU(N)$ subgroup obey the algebra $[T^{a}, T^{b}]=\imath
f^{abc}T^{c},$ where $f^{abc}$ are totally antisymmetric structure
constants of the gauge group. One can also define an
anticommutator as $\{T^{a}, T^{b}\}=\frac{1}{N}\delta^{ab}+
d^{abc}T^{c}$ with symmetric structure constants $d^{abc}$. The
completeness relation is written in the form (here, as usual, the
repeated indices are summed over)
$$T^{a}_{\alpha\beta}T^{a}_{\gamma\delta}=\frac{1}{2}\delta_{\alpha\delta}\delta_{\beta\gamma}-\frac{1}{2N}\delta_{\alpha\beta}\delta_{\gamma\delta}$$
which can be used to derive the following useful identities:
\begin{equation}\label{thirtyone}
T^{a}_{\alpha\beta}T^{a}_{\beta\gamma}=\frac{N^{2}-1}{2N}\delta_{\alpha\gamma},~~
T^{A}_{\alpha\beta}T^{A}_{\beta\gamma}=\frac{N}{2}\delta_{\alpha\gamma}.
\end{equation}
The gauge field kinetic operator (\ref{one}) is  represented in
components as
\begin{equation}\label{thirtytwo}
(D^{\xi}_{\mu\nu})_{\alpha\beta}=-\Bigl[\delta_{\mu\nu}\nabla^{2}+(\frac{1}{\xi}-1)\nabla_{\mu}
 \nabla_{\nu}
 +2[\widehat{F}_{\mu\nu},~\cdot~]_{\star}\Bigr]_{\alpha\beta}.
\end{equation}
 It can
be easily seen that the relations (\ref{nineb})-(\ref{fourteen})
remain unchanged with the only difference that now one looks at
them as matrix relations. Similarly, to define heat kernels for
the operators $(D_{0})_{\alpha\beta}$ and
$(D^{\xi}_{\mu\nu})_{\alpha\beta}$ one can introduce two abstract
Hilbert spaces spanned by basis vectors $|x, A \rangle$ and $|\mu,
x, A \rangle$, respectively, which satisfy the orthonormality
conditions $$\langle A, x| x', B \rangle
=\delta(x,x')\delta^{AB},~~~~~~~~~$$
 $$ \langle A, x, \mu|\nu,x', B \rangle =\delta_{\mu\nu}\delta(x,x')\delta^{AB}.$$
Then the heat kernels
 for the operators $D_{0}$ and
$D^{\xi}_{\mu\nu}$ are defined  by
\begin{eqnarray}\label{thirtythree}
K_{0}(x, x'; t)= ( x|\exp[-t \widehat{D}_{0}]|x' ),~~~~~~~
\nonumber
\\ K^{\xi}_{\mu\nu}(x, x'; t)= ( x, \mu|\exp[-t
\widehat{D}^{\xi}]|\nu,x' ),
\end{eqnarray}
where we denote
 $(x | =T^{A} \langle A, x |:=\sum_{A=1}^{N^{2}}T^{A} \langle A, x |$ and $( x, \mu|=T^{A}\langle  A, x, \mu
|$. It is straightforward to verify that the formulae
(\ref{eighteenprime})  - (\ref{twentyfour}) remain unchanged as
well (one should only replace $\langle x |$ or $\langle x, \mu |$
by $(x |$ or $( x, \mu|$, where it is necessary). As an example,
consider the planar contribution to the quantity
(\ref{twentyfour}). In the $U(N)$ case it reads
\begin{eqnarray}\label{thirtyfour}
 \int_{R^{n}}dx ~  \mathrm{tr_{N}}~ \Bigl[\{ \nabla^{L}_{\mu}
\nabla^{L}_{\mu}e^{-t D^{L}_{0}}+\nabla^{R}_{\mu}
\nabla^{R}_{\mu}e^{-t D^{R}_{0}}-\partial^{2}e^{t\partial^{2}}\}(
x~|~ x') \Bigr]_{x=x'}.
\end{eqnarray}
With the help of (\ref{thirtyone}) and the orthonormality
condition one has $$( x~|~ x')_{\alpha\beta}=T^{A}_{\alpha\gamma}
\langle A, x |x', B \rangle T^{B}_{\gamma
\beta}=T^{A}_{\alpha\gamma}T^{A}_{\gamma \beta}\langle x|x'
\rangle=\frac{N}{2}\delta_{\alpha\beta}\langle x|x' \rangle.$$ By
making use of the plane-wave basis and applying the calculating
technique of the preceding section one obtains:
\begin{eqnarray}\label{thirtyfive}
  \widetilde{a}^{planar}_{2}(\nabla^{2}, D_0)&=&\frac{1}{(4 \pi)^{\frac{n}{2}}}\frac{N}{2}\int_{R^{n}}dx
  \frac{4-n}{24}~\mathrm{tr_{N}}~F_{\mu\nu}\star
F_{\mu\nu},\\
 \widetilde{a}^{planar}_{4}(\nabla^{2}, D_0)&=&\frac{1}{(4
 \pi)^{\frac{n}{2}}}\frac{N}{2}\int_{R^{n}}dx\frac{n-6}{360}~\mathrm{tr_{N}}~(6 F_{\mu\nu}\star
F_{\nu\lambda}\star F_{\lambda\mu}\nonumber\\
&~&~~~~~~~~+2\nabla_{\mu}F_{\nu\lambda}\star\nabla_{\mu}F_{\nu\lambda}-\nabla_{\mu}F_{\mu\lambda}\star\nabla_{\nu}F_{\nu\lambda})\nonumber.
\end{eqnarray}
where $\mathrm{tr_{N}}$ means trace over internal indices. The
mixed terms are treated in a similar way. One gets, in particular,
\begin{eqnarray}\label{thirtysix}
  \widetilde{a}^{mixed}_{n}(\nabla^{2}, D_0)=-2(det~\theta)^{-1}(2\pi)^{-n}\int_{R^{n}}dx
\int_{R^{n}}dy  ~\mathrm{tr_{N}}~B_{\mu}(x)T^{D} B_{\mu}(y)T^{D}.
\end{eqnarray}
Next,
\begin{eqnarray*}\label{}
 \mathrm{tr_{N}}~T^{A}T^{D}T^{C}T^{D}= \mathrm{tr_{N}}~\left(\frac{1}{2N}T^{A}T^{C}+T^{A}T^{d}T^{C}T^{d}\right)~~~~~~~~~~~~~~~~\\
 =\mathrm{tr_{N}}~\left(\frac{1}{2N}T^{A}T^{C}\right)+\frac{1}{2}T^{A}_{\alpha
\alpha}T^{C}_{\beta \beta} -\frac{1}{2N}T^{A}_{\alpha
\beta}T^{C}_{\beta
\alpha}=\frac{1}{2}\mathrm{tr_{N}}T^{A}~\mathrm{tr_{N}}T^{C},
\end{eqnarray*}
where we used the completeness relation for the generators of the
$SU(N)$ group. Hence one arrived at the following expression for
$\widetilde{a}^{mixed}_{n}$:
\begin{eqnarray}\label{thirtyseven}
  \widetilde{a}^{mixed}_{n}(\nabla^{2}, D_0)=-\frac{(det~\theta)^{-1}}{(2\pi)^{n}}\int_{R^{n}}dx
\int_{R^{n}}dy  ~\mathrm{tr_{N}}B_{\mu}(x) ~\mathrm{tr_{N}}
B_{\mu}(y)\nonumber\\
=-\frac{(det~\theta)^{-1}}{2(2\pi)^{n}}\int_{R^{n}}dx
\int_{R^{n}}dy  ~B^{0}_{\mu}(x) ~ B^{0}_{\mu}(y).
\end{eqnarray}
This expression is manifestly gauge invariant and depends only
upon zeroth component of the gauge potential.  According to the
formula (\ref{twentyeight}), the planar heat kernel coefficients
for the operator (\ref{thirtytwo}) are given by
\begin{eqnarray*}
   a^{planar}_{4}=\frac{1}{(4 \pi)^{\frac{n}{2}}}~\frac{N}{2}\int_{R^{n}}dx
  \left(\frac{n}{6}-1+\frac{1}{12}(1-\xi^{\frac{n-4}{2}})\right)~\mathrm{tr_{N}}~F_{\mu\nu}\star
F_{\mu\nu},
\end{eqnarray*}
\begin{eqnarray}\label{thirtyeight}
 a^{planar}_{6}=\frac{1}{(4
\pi)^{\frac{n}{2}}}\frac{1}{360}~\frac{N}{2}~\int_{R^{n}}dx
~\mathrm{tr_{N}}~\{120
   F_{\mu\nu}\star
F_{\nu\lambda}\star F_{\lambda\mu}~~~~~~~~~~~~~~\\
 ~~~-60F_{\mu\nu}\star\nabla^{2}
F_{\mu\nu}\nonumber - 2[n+1-\xi^{\frac{n-6}{2}}](6 F_{\mu\nu}\star
F_{\nu\lambda}\star F_{\lambda\mu}\nonumber\\
~~~~~~~~~~~~~~~~~~~~~~~~~~~~~~~+2\nabla_{\mu}F_{\nu\lambda}\star\nabla_{\mu}F_{\nu\lambda}
-\nabla_{\mu}F_{\mu\lambda}\star\nabla_{\nu}F_{\nu\lambda})\}.\nonumber
\end{eqnarray}
The first non-zero mixed coefficient is written as
\begin{eqnarray}\label{thirtynine}
  a^{mixed}_{n+2}=\{2(n-1)+\xi^{-1}\}\frac{(det~\theta)^{-1}}{2(2\pi)^{n}}\int_{R^{n}}dx
 \int_{R^{n}}dy  ~B^{0}_{\mu}(x) ~ B^{0}_{\mu}(y).
\end{eqnarray}

At the end of this section, let us make a few remarks about the
obtained results. We consider the particular case of dimension
$n=4$ for the purpose of definiteness. First, it is seen from
(\ref{thirtyeight}) that the fourth heat kernel coefficient do not
depend upon the gauge fixing parameter $\xi$. Thus, the one-loop
$\beta$-function is a gauge-fixing independent object as it is in
the commutative Yang-Mills theory (see, for instance, Refs.
\cite{Barvinsky:1985bs}, \cite{Guendelman:1993ke}).

Second, in the case of a non-degenerate $\theta$ matrix the
one-loop renormalization of the theory is not affected by the
mixed coefficients. Moreover, they are completely determined by
$U(1)$ sector of the model.  In the diagrammatic approach this
implies the known fact that non-planar one-loop $U(N)$ diagrams
contribute only to the $U(1)$ part of the theory
\cite{Minwalla:1999we}, \cite{Armoni:2000xi}. As it was mentioned,
such coefficients are responsible for the UV/IR mixing phenomenon
\cite{Gayral:2004ww}, \cite{Vassilevich:2005vk}.

Third, in the case of a degenerate deformation parameter the first
non-trivial mixed contribution  appears already in
$a_{4}$-coefficient (see also the recent paper
\cite{Gayral:2006vd}). To see this let us examine the space-like
noncommutativity when components $\theta^{0i}$, $i=1,2,3$ are
equal to zero. For convenience, we adopt the same conventions as
in  Ref. \cite{Gayral:2004cu} (see Appendix B for details). Then
after simple manipulations one gets
\begin{eqnarray}\label{fourty}
 a^{mixed}_{4}=\frac{(det\theta_{2})^{-1}}{32\pi^{3}}(8+\ln{\xi})\int_{R^{2}}d\widetilde{x}~\int_{R^{2} \times R^{2}}
 d\overline{x}~d\overline{y}~\sum_{i=2,3}B^{0}_{i}(\widetilde{x},\overline{x}) ~
 B^{0}_{i}(\widetilde{x},\overline{y}),
\end{eqnarray}
where tensor $\theta_{2}$ corresponds to the  $i=2,3$ plane. Note
that, contrary to its planar counterpart, this coefficient itself
is  dependent on the gauge-fixing parameter. Next, it can be
easily shown that a  non-planar divergent part of the one-loop
effective action for the $U(1)$  sector of the model is presented
by\footnote{For the sake of brevity  we do not consider the ghost
contribution here. We remark only that this contribution does not
change our conclusion.}
\begin{eqnarray*}\label{}
 \Gamma^{div.}_{NP}[B^{0}]= -\frac{\mu^{2\epsilon}}{32\pi^{3}det \theta_{2}}(8+\ln{\xi})\int_{0}^{\infty}\frac{dt}{t^{\epsilon}}\int_{R^{2}}d\widetilde{x}~\int_{R^{2} \times R^{2}}
 d\overline{x}~d\overline{y}~~~~~~~~~~~~~~~~~~~~~~~~~\nonumber\\
 \times \sum_{i=2,3}B^{0}_{i}(\widetilde{x},\overline{x}) ~
 B^{0}_{i}(\widetilde{x},\overline{y})\exp[-\frac{t(\overline{x}-\overline{y})^{2}}{det \theta_{2}}],
\end{eqnarray*}
which gives the non-local, singular (as $\epsilon \rightarrow 0 $)
and, in addition, gauge-fixing dependent  contribution to the 1PI
2-point Green function of the form \cite{Gayral:2004cu}
\begin{eqnarray*}\label{}
   \Gamma^{div.}_{U(1)}(x_{1}-x_{2})= -\frac{\mu^{2\epsilon}}{16\pi^{3}det \theta_{2}}(8+\ln{\xi})\Gamma(\epsilon)\delta^{2}(\widetilde{x}_{1}-\widetilde{x}_{2})
   \left(\frac{(\overline{x}_{1}-\overline{x}_{2})^{2}}{det \theta_{2}}\right)^{-\epsilon}.
\end{eqnarray*}
 Hence we come to the
conclusion that the renormalization properties of  NC $U(N)$
theory are actually ruined by its $U(1)$ sector in the degenerate
case. It looks rather surprising but even this crucial drawback of
the space-like noncommutative models may be bypassed in the
context of
 so-called twisted gauge transformations considered in the next
section.

\section{Remarks on twisted gauge symmetries}
Originally twisted symmetries appeared in  NCQFT as an attempt to
resolve the problem of the lack of  Lorenz invariance which is
known to be an attribute of such theories. In particular, it was
realized that it is possible to formulate a Lorenz-invariant QFT
on a deformed manifold if the coproduct of the universal envelope
of the Poincare algebra is twisted in  such a   way that it is
compatible with Moyal product \cite{Chaichian:2004hh}. Since the
twist affects solely the action of the Poincare generators in the
tensor product of  Poincare group representations,  but not the
algebra of the generators itself,  NCQFT with the twisted Poincare
symmetry possesses the same representation content as the usual
commutative Lorenz-invariant theory that in turn validates the
consideration of many other aspects of NC field theories like
unitarity and causality. Other issues on this fast-developing
topic can be found, for instance, in \cite{common},
\cite{Tureanu:2006pb}, \cite{Alvarez-Gaume:2006bn}.

Another essential peculiarity  of  NCQFT consists in the
well-known fact that not every gauge group is closed  with respect
to the star-gauge transformations which underlie the standard
approach to NC gauge theories. Recently it was proposed an idea
that not only Poincare but also internal symmetry can be twisted
\cite{Vassilevich:2006tc}, \cite{Aschieri:2006ye} (see also
\cite{Chaichian:2006we} for alternative point of view).  In this
approach the transformation law of primary fields is left
undeformed while the comultiplication is twisted leading to the
deformation of the Leibniz rule. In particular,  this implies that
there is a certain amount of freedom in the choice of gauge group
for construction of NC gauge field models. Namely, for a gauge
group with generators $\tau_{N}$ in some representation and gauge
parameter $\sigma=\sigma^{N}\tau^{N}$, the transformation law for
the gauge potential $B_{\mu}=B^{N}_{\mu}\tau^{N}$ is taken to be
the usual one (and thus the gauge group is automatically close),
i.e.
\begin{equation}\label{fourtyone}
\delta_{\sigma}B_{\mu}=\partial_{\mu}~\sigma+[\sigma,~B_{\mu}].
\end{equation}
 The coproduct
$\bigtriangleup_{0}(\delta_{\sigma})=\delta_{\sigma}\otimes
1+1\otimes \delta_{\sigma} $ is deformed as
$$\bigtriangleup_{\mathcal{F}}(\delta_{\sigma})=\mathcal{F}^{-1}\bigtriangleup_{0}(\delta_{\sigma})\mathcal{F},$$
where
$\mathcal{F}=\exp[\frac{i}{2}\theta^{\alpha\beta}\partial_{\alpha}\otimes
\partial_{\beta}]$ is the twist operator. Consequently, the action of the gauge
transformations on the star-product of two gauge fields (Leibniz
rule) reads \cite{Vassilevich:2006tc}, \cite{Aschieri:2006ye}
\begin{eqnarray*}\label{}
\delta_{\sigma}(B_{\mu}\ast B_{\nu})=\mu_{\star} \circ
\bigtriangleup_{\mathcal{F}}(\delta_{\sigma})(B_{\mu}\otimes
B_{\nu})=\partial_{\mu}\sigma \cdot B_{\nu}+\partial_{\nu}\sigma
\cdot B_{\mu}+[\sigma,~ B_{\mu}\ast B_{\nu}],
\end{eqnarray*}
which clearly differs from the rule $(\delta_{\sigma}B_{\mu})\ast
B_{\nu}+B_{\mu}\ast(\delta_{\sigma}B_{\nu})$ that appears in the
star-gauge transformation approach. In the formula above
$\mu_{\star}$ denotes a map that maps tensor product of two
functions $f$ and $g$ to the space of functions with star-product:
$\mu_{\star}\{f \otimes g\}= f \star g$. Similarly, one can show
that the operator (\ref{one}) transforms covariantly under the
transformations (\ref{fourtyone}),
\begin{equation}\label{fourtytwo}
 \delta_{\sigma}D^{\xi}_{\mu\nu}=[\sigma,~
 D^{\xi}_{\mu\nu}],
\end{equation}
that is, the heat trace (\ref{two}) is the twisted-gauge invariant
object. It is straightforward to prove that all presented in the
sections 4-5 results are valid in the twisted approach as well.
However,  the twisted principle allows for the existence of other
gauge symmetries, like, for instance, $SU(N)$ gauge group,  on its
own right. As we saw in the previous section, the mixed
contribution  to the heat kernel involves only the $U(1)$ sector
of the model  while the $SU(N)$ sector has no pathological terms
at all. One might expect, therefore, that it is possible to get
rid of these terms just by restricting oneself to the
consideration of the $SU(N)$ gauge group as a particular symmetry
group of the model, solving in such a way the problem of the
renormalizability
 of space-like NC non-Abelian gauge theories and even of the presence  of UV/IR
mixing phenomenon\footnote{There was a claim in the literature
about the absence  of UV/IR mixing in NC models considered
 in the realm of the
 twisted Poincare symmetry; this was, in fact, a consequence of the deformed
 commutation relations imposed on field operators in quantization
 of a model.
  Recently it was argued, however, that within a canonical quantization procedure (with usual commutation relations) the  presence or absence of the
 mixing phenomenon is dependent  on a particular choice of the interaction
 term of the action  and not
 on the twisted structure of the model \cite{Tureanu:2006pb}.},
  at one-loop level at least. This is
not true, however, since, as it is argued in Ref.
\cite{Aschieri:2006ye}\footnote{For further discussion on the
topic see also  Ref. \cite{Alvarez-Gaume:2006bn}.}, the
consistency of the equation of motion for the twisted YM fields
requires in this case to add additional vector potentials into the
action. These auxiliary fields would bring back mixed terms into
the heat trace and, consequently, dangerous non-planar
contributions into the one-loop effective action although the
$SU(N)$ Lie-algebra-valued field dynamics will remain the same as
that of the usual YM fields leading, in particular, to the same
one-loop (gauge-fixing independent) counterterms and
$\beta$-function. Nevertheless one should not exclude the
possibility of the existence of other twisted invariant theories
(apart from the $U(N)$ gauge model considered in the paper) within
which the problem of inconsistency  can be resolved.

Of course, there is an alternative way to formulate a NC $SU(N)$
model which is based on the Seiberg-Witten map between the
commutative and noncommutative theories \cite{seiberg}. Note,
however, that the non-planar heat kernel coefficients have
singular terms in $\theta$ expansion when $\theta \rightarrow 0$.

\section{Summary}
 In
this paper we calculated the heat trace asymptotic for the
non-minimal gauge field kinetic operator on the Moyal plane within
background field formalism. We found, in particular, that the
planar part of the heat trace expansion, although being dependent
on the gauge-fixing parameter in general,
 leads to
gauge-fixing independent counterterms of the model which is in
accordance with well-known results of conventional gauge field
theories (see, for instance, \cite{Barvinsky:1985bs},
\cite{Guendelman:1993ke}). In the case of a degenerate deformation
parameter this picture, however, is spoiled by the non-planar part
that develops  non-local and gauge-fixing dependent singularities
of the effective action; the phenomenon was discovered at first in
Ref. \cite{Gayral:2004cu} (see also the recent paper
\cite{Gayral:2006vd}) for the space-like noncommutative scalar
$\lambda \varphi^{4}$
 and gauge $U(1)$ models. The latter observation concerning space-like noncommutativity
  is especially unpleasant since, first, this
makes it rather difficult to formulate a NC field theory on
odd-dimensional manifolds (though it is possible to avoid such
difficulty in some particular physical systems, such as  field
theories in a thermal medium considered in the imaginary time
formalism\footnote{On the other hand, just for the same reason,
one may get into trouble if he will try to consider NC finite
temperature theories on even dimensional space-time.}, further
discussion of this point can be found in
\cite{Vassilevich:2005vk}) and, second, just space-time
noncommutativity  leads to the well-known problems with unitarity
and causality \cite{common2}.

 In the  case
 of the $U(N)$ gauge symmetry the non-planar contribution
 to the heat trace expansion is shown to be determined only by the $U(1)$ sector of the
 model  that is in agreement with previous results in the
literature coming from the diagrammatic approach
\cite{Minwalla:1999we}, \cite{Armoni:2000xi}.

\subsection*{Acknowledgments}

\hspace{\parindent} The author is sincerely grateful to Dmitri
Vassilevich for numerous helpful discussions and comments and, in
particular, for bringing his attention to the papers by Endo.

\appendix
\section{Calculation of the planar heat kernel
         coefficients on Moyal plane}
In this section we present the evaluation  of the asymptotic
expansion for the quantity $$Tr_{L^{2}}(Q e^{-t D})~,$$ where $Q$
is  a a second order star-differential operator with scalar
leading symbol\footnote{That is, with the leading symbol of the
form $h^{\mu\nu}\xi_{\mu}\xi_{\nu}$, where $h^{\mu\nu}$ is a
(constant) symmetric tensor of (2,0) type.} and $D$ is a (minimal)
star-Laplace type operator. We assume that the operator $Q e^{-t
D}:=Q^{L} e^{-t D^{L}}$ contains only left Moyal multiplications.
It corresponds, in particular, to the first term  in the RHS of
Eq. (\ref{twentyfour}). The operator $D^{L}$ can be represented in
the canonical form as
$$D^{L}=-(\delta^{\mu\nu}\nabla_{\mu}^{L}\nabla_{\nu}^{L}+L(E)),~$$
$$\nabla_{\mu}^{L}=\partial_{\mu}+\omega^{L}_{\mu},~~\omega_{\mu}^{L}:=L(\omega_{\mu}),$$
with $\omega_{\mu}$ and $E$ being bundle connection and bundle
endomorphism, respectively.
 To introduce the operator $Q^{L}$ let us
determine, following \cite{Branson:1997Bn},  variations of the
(flat) metric $g(\varepsilon)=\delta+\varepsilon q_{2}$ and of the
connection $\omega_{\mu}(\varepsilon)=\omega_{\mu}+\varepsilon
q_{1}$, where $q_{2~\mu\nu}$ is a symmetric constant\footnote{This
is necessary to keep the metric flat. Otherwise the variation of
the metric would affect the Moyal product as well.} tensor and
$q_{1}^{\mu}$ is an endomorphism valued 1-tensor. Then we define
$Q^{L}$ by $$Q^{L}=Q^{L}_{2}+Q^{L}_{1}+L(Q_{0}),$$ where
$Q^{L}_{2}=\partial_{\varepsilon}D^{L}(g(\varepsilon),\omega_{\mu}^{L},E)|_{\varepsilon=0}$,
$Q^{L}_{1}=\partial_{\varepsilon}D^{L}(\delta,\omega_{\mu}^{L}(\varepsilon),E)|_{\varepsilon=0}$
and $Q_{0}$ is an endomorphism of bundle. Making use of the
explicit form of $D^{L}$ it is easy to derive for the operators
$Q^{L}_{1,2}$ (cf. Lemma 2.4 in \cite{Branson:1997Bn}):
$$Q^{L}_{2}=-\partial_{\varepsilon}(g(\varepsilon))^{\mu\nu}|_{\varepsilon=0}\nabla_{\mu}^{L}\nabla_{\nu}^{L}=q_{2~\mu\nu}\nabla_{\mu}^{L}\nabla_{\nu}^{L},$$
 $$Q^{L}_{1}=
-\partial_{\varepsilon}\left(\nabla^{L}_{\mu}(\varepsilon)\nabla^{L}_{\mu}(\varepsilon)\right)|_{\varepsilon=0}=-\{\nabla_{\mu}^{L},q_{1}^{\mu}\},$$
where
$\nabla^{L}_{\mu}(\varepsilon)=\partial_{\mu}+L(\omega_{\mu}(\varepsilon))$
and $\{~,~\}$ stands for the usual operator anticommutator. Note
also that
$\partial_{\varepsilon}(g(\varepsilon))^{\mu\nu}|_{\varepsilon=0}=-q_{2~\mu\nu}.$

Now we are interesting in  $t \rightarrow 0$ asymptotic of the
trace $Tr_{L^{2}}(Q^{L} e^{-t D^{L}})$. It is written in the form
(cf. expr. (\ref{twentytwo}))
\begin{equation}\label{hundred}
 Tr_{L^{2}}(Q^{L} e^{-t D^{L}})\simeq
\sum_{k=-2}^{\infty}t^{(k-n)/2}\widetilde{a}^{L}_k(Q^{L}, D^{L}),
\end{equation}
where the  coefficients $\widetilde{a}^{L}$ can be decomposed as
\begin{equation}\label{hundredone}
 \widetilde{a}^{L}_k(Q^{L}, D^{L})=\widetilde{a}^{L}_k(Q^{L}_{2},
D^{L})+\widetilde{a}^{L}_k(Q^{L}_{1},
D^{L})+\widetilde{a}^{L}_k(Q^{L}_{0}, D^{L}).
\end{equation}
 Hence it is
sufficient to calculate the heat trace coefficients for each
operator $Q^{L}_{i}$. Coefficients $\widetilde{a}^{L}_k(Q^{L}_{0},
D^{L})$ are calculated in Refs. \cite{Vassilevich:2003yz},
\cite{Gayral:2004ww}:
\begin{eqnarray}\label{hundredtwo}
\widetilde{a}^{L}_2(Q^{L}_{0}, D^{L})&=&\frac{1}{(4
\pi)^{\frac{n}{2}}}\int_{R^{n}}dx~ \mathrm{tr}~ Q_{0}\star E,
\nonumber\\ \widetilde{a}^{L}_4(Q^{L}_{0}, D^{L})&=&\frac{1}{(4
\pi)^{\frac{n}{2}}}\int_{R^{n}}dx ~ \mathrm{tr}~ Q_{0}\star
\left(\frac{1}{2}E\star E
+\frac{1}{6}\nabla^{2}E+\frac{1}{12}F_{\mu\nu}\star
F_{\mu\nu}\right), \nonumber\\
\widetilde{a}^{L}_6(Q^{L}_{0},D^{L})&=&\frac{1}{(4
\pi)^{\frac{n}{2}}}\int_{R^{n}}dx ~ \mathrm{tr}~ Q_{0}\star \{
\frac{1}{6}E \star E \star E + \frac{1}{12}E \star \nabla^{2}E
\nonumber\\&~& +\frac{1}{12}E \star F_{\mu\nu}\star
F_{\mu\nu}+\frac{1}{60}\nabla^{2}\nabla^{2}E
\\
&~&-\frac{1}{180}\left(6 F_{\mu\nu}\star F_{\nu\lambda}\star
F_{\lambda\mu}+2\nabla_{\mu}F_{\nu\lambda}\star\nabla_{\mu}F_{\nu\lambda}-\nabla_{\mu}F_{\mu\lambda}\star\nabla_{\nu}F_{\nu\lambda}\right)\}\nonumber,
\end{eqnarray}
where
$\nabla_{\mu}=\partial_{\mu}+L(\omega_{\mu})-R(\omega_{\mu})$ and
$F_{\mu\nu}=\partial_{\mu}\omega_{\nu}-\partial_{\nu}\omega_{\mu}+[\omega_{\mu},\omega_{\nu}]_{\star}$
is the curvature of the connection $\omega$. To evaluate the
remaining coefficients one can use the method of Ref.
\cite{Branson:1997Bn}. Consider an 1-parameter family of
star-Laplacians $D(\rho)$. Then the following relation for the
heat trace coefficients  holds:
$$\widetilde{a}^{L}_n(\partial_{\rho}D^{L}(\rho),
D^{L}(\rho))=-\partial_{\rho}\widetilde{a}^{L}_{n+2}(1,
D^{L}(\rho)).$$ In particular, we apply this formula to the
calculation of the coefficients in (\ref{hundredone}). For the
operator $Q^{L}_{2}$ we have:
\begin{eqnarray}\label{hundredthree}
 \widetilde{a}^{L}_n(\partial_{\varepsilon}D^{L}(\widehat{\delta}(\varepsilon),\omega_{\mu}^{L},E),
D^{L}(\widehat{\delta}(\varepsilon),\omega_{\mu}^{L},E))|_{\varepsilon=0}=~~~~~~~~~~~~~~~~~~\nonumber\\
=\widetilde{a}^{L}_n(Q^{L}_{2},
D^{L})=-\partial_{\varepsilon}\widetilde{a}^{L}_{n+2}(1,
D^{L}(\widehat{\delta}(\varepsilon),\omega_{\mu}^{L},E))|_{\varepsilon=0}.
\end{eqnarray}
Taking  the variation of the volume form,
$$\partial_{\varepsilon}|_{\varepsilon=0}d_{\varepsilon}\mathrm{vol}(x)
= \frac{1}{2}q_{2~\mu\mu}d ~\mathrm{vol}(x),$$ into account one
gets (see Theorem 3.3 of Ref. \cite{Branson:1997Bn})
\begin{eqnarray}\label{hundredfour}
\widetilde{a}^{L}_{-2,0}(Q^{L}_{2},
D^{L})=-\frac{1}{2}\widetilde{a}^{L}_{0,2}(q_{2~\mu\mu}, D^{L}),
~~~~~~~~~~~~~~~~~~~~~~~~~~~~~~~~~~~~~~~~~~~~\nonumber\\
  \widetilde{a}^{L}_{2}(Q^{L}_{2},
D^{L})=-\frac{1}{2}\widetilde{a}^{L}_{4}(q_{2~\mu\mu},
D^{L})~~~~~~~~~~~~~~~~~~~~~~~~~~~~~~~~~~~~~~~~~~~~~~~~~~\\
+\frac{1}{(4 \pi)^{\frac{n}{2}}}\int_{R^{n}}dx
~\frac{1}{6}~q_{2~\mu\nu}~ \mathrm{tr}~
\left(\nabla_{\mu}\nabla_{\nu}E+F_{\mu\lambda}\star
F_{\nu\lambda}\right) ,~~~~~~~~ \nonumber\\
 \widetilde{a}^{L}_{4}(Q^{L}_{2},
D^{L})=-\frac{1}{2}\widetilde{a}^{L}_{6}(q_{2~\mu\mu}, D^{L})
+\frac{1}{(4 \pi)^{\frac{n}{2}}}\int_{R^{n}}dx q_{2~\mu\nu}
\mathrm{tr}~\{\frac{1}{12}E \star
\nabla_{\mu}\nabla_{\nu}E\nonumber\\
 +\frac{1}{6}E \star
F_{\mu\lambda}\star
F_{\nu\lambda}+\frac{1}{30}\nabla_{\mu}\nabla_{\nu}\nabla^{2}E
-\frac{1}{180}(18 F_{\mu\rho}\star F_{\rho\lambda}\star
F_{\lambda\nu}\nonumber\\
-2\nabla_{\mu}F_{\rho\lambda}\star\nabla_{\nu}F_{\rho\lambda}
-4\nabla_{\rho}F_{\mu\lambda}\star\nabla_{\rho}F_{\nu\lambda}~~\nonumber\\
+2\nabla_{\mu}F_{\nu\lambda}\star\nabla_{\rho}F_{\rho\lambda}+\nabla_{\lambda}F_{\lambda\mu}\star\nabla_{\rho}F_{\rho\nu})\}\nonumber.
\end{eqnarray}
In exact analogy to the described above procedure one can
calculate planar coefficients for operators of the type $Q^{R}
e^{-t D^{R}}$, where
$D^{R}=-\delta^{\mu\nu}\nabla_{\mu}^{R}\nabla_{\nu}^{R}-R(E)$ and
$Q^{R}$ is an operator with "scalar leading symbol" containing
only right Moyal multiplications. In particular, by putting $E=0$,
$Q_0=1$, $q_{1}=0$, $q_{2}=I$ and $\omega_{\mu}=B_{\mu}$ one
reproduces formulae (\ref{thirtyfive}).

\section{Calculation of the mixed heat kernel coefficients [11]}
In the computation of the mixed contribution to the heat trace
(\ref{twentytwo}), (\ref{twentyeight}) one encounters  the
following typical integral :
\begin{equation}\label{twohundred}
 T(l,r)=\int_{R^{n}}dx \int \frac{d^{n}k}{(2\pi)^{n}} e^{-t k^{2}}
 e^{-\imath k x}L(l(x))\circ R(r(x)) e^{\imath k x},
\end{equation}
where $l(x)$ and $r(x)$ are some smooth rapidly decreasing on
$R^{n}$ functions. To evaluate it we proceed as follows. First, we
note that (in the case of a non-degenerate noncommutativity
parameter) the star-product can be represented by\footnote{This is
the so-called Rieffel representation of the star-product.} (see
also \cite{Gayral:2004cu}):
\begin{equation}\label{twohundredone}
  l \star r(x)=\frac{1}{(2 \pi)^{n}}\int_{R^{n} \times R^{n}}du~ dv
  e^{-\imath uv}~
l(x-\frac{1}{2}\theta u)~r(x+v),
  \end{equation}
where $\theta$ is any real skewsymmetric $n \times n$ matrix  and
$\theta u $ means (in components) $ (\theta u
)^{\mu}=\theta^{\mu}_{~\nu} u^{\nu}$. Hence, for the integrand in
(\ref{twohundred}) one gets:
\begin{eqnarray}\label{twohundredtwo}
  L(l(x))\circ R(r(x)) e^{\imath k x}:=l(x)\star e^{\imath k x} \star
  r(x)
  =\frac{1}{(2 \pi)^{2n}}\int_{R^{n} \times R^{n}}du~ dv ~e^{-\imath uv}~~~~~~~~~~~~~~\nonumber\\
  \times \int_{R^{n} \times R^{n}}du'~
  dv'~e^{-\imath u'v'}~e^{\imath u'v}~\exp[\imath k (x+v-\frac{1}{2}\theta u')]~l(x-\frac{1}{2}\theta
  u)~r(x+v').
\end{eqnarray}
Next, the integration over variables $v$ and $u'$ is
straightforward and (\ref{twohundred}) is rewritten as:
\begin{eqnarray}\label{twohundredthree}
 T(l,r)=\int_{R^{n}}dx \int \frac{d^{n}k}{(2\pi)^{n}} e^{-t k^{2}}
\int_{R^{n} \times R^{n}}\frac{du~ dv'}{(2\pi)^{n}}~e^{\imath
v'(k-u)} e^{-\frac{\imath}{2} k\theta u}~l(x-\frac{1}{2}\theta
  u)~r(x+v'),
\end{eqnarray}
which after a suitable change of integration variables, $(x,v')
\rightarrow (y, z)$ with $y=x-\frac{1}{2}\theta
  u$ and $z=x+v'$,  can be cast into the form
\begin{eqnarray*}
 T(l,r)=\int \frac{d^{n}k}{(2\pi)^{n}} e^{-t k^{2}}\int_{R^{n}}\frac{du}{(2\pi)^{n}}e^{-\imath k\theta u}
\int_{R^{n} \times R^{n}}dy~ dz~e^{-\imath (y-z)(k-u)} ~l(y)~r(z).
\end{eqnarray*}
Finally,  the integrals over $u$ and $k$ are carried out trivially
and  we obtain:
\begin{eqnarray}\label{twohundredfour}
 T(l,r)=\frac{(det\theta)^{-1}}{(2 \pi)^{n}}\int_{R^{n} \times R^{n}}dy~
 dz
\exp[-t (\theta
\theta^{T})^{-1}_{\mu\nu}(y-z)^{\mu}(y-z)^{\nu}]~l(y)~r(z).
\end{eqnarray}
Note that to derive the mixed heat kernel coefficients one has to
expand the exponential in (\ref{twohundredfour}) in series over
 parameter $t$ as well. In this way one reproduces the result
of Ref. \cite{Vassilevich:2005vk} (cf. formulae (31)~- (33)
therein).

It is instructive to consider the case of a degenerate deformation
parameter. Without loss of generality we assume that the first
$n-m$, $n\geq m$, coordinates commute. The whole manifold is split
into two subspaces with coordinates denoted by $\widetilde{x}$ for
commutative submanifold and by $\overline{x}$ for noncommutative
one, i.e. the coordinate of a point in $R^{n}$ is written in our
conventions as
$x=(\widetilde{x},\overline{x})$\footnote{Similarly, in momentum
space one writes $k=(\widetilde{k},\overline{k})$.}. Then
noncommutativity of the model is encoded in a (non-degenerate)
skewsymmetric $m \times m$ matrix $\theta_{m}$. The star-product
of two functions $l(x)$ and $r(x)$ is now given by
\begin{equation}\label{twohundredfive}
  l \star r(x)=\frac{1}{(2 \pi)^{m}}\int_{R^{m} \times R^{m}}d\overline{u}~ d\overline{v}
  e^{-\imath \overline{u}~\overline{v}}~
l(\widetilde{x}, \overline{x}-\frac{1}{2}\theta_{m}
\overline{u})~r(\widetilde{x},\overline{x}+\overline{v}).
  \end{equation}
The quantity (\ref{twohundred}) is evaluated completely in the
same manner as it was done for the non-degenerate case above. The
result reads
\begin{eqnarray}\label{twohundredsix}
 T(l,r)=\frac{(det\theta_{m})^{-1}}{2^{n}\pi^{\frac{n+m}{2}}}t^{\frac{m-n}{2}}\int_{R^{n-m}}d\widetilde{x}~\int_{R^{m} \times R^{m}}
 d\overline{x}~d\overline{y}~~~~~~~~~~~~~~~~~~~~\nonumber\\
\times \exp[-t (\theta_{m}
\theta_{m}^{T})^{-1}_{\mu\nu}(\overline{x}-\overline{y})^{\mu}(\overline{x}-\overline{y})^{\nu}]~
l(\widetilde{x},\overline{x})~r(\widetilde{x},\overline{y}).
\end{eqnarray}
From this expression one can see that in the case of the
degenerate deformation parameter the first nontrivial  mixed
contribution will appear earlier than in $a^{mixed}_{n+2}$. Such
terms can produce non-local singularities in  the effective action
of a space-like noncommutative theory \cite{Gayral:2004cu}.

\vspace{2cm}

\end{document}